\documentclass[a4paper]{article} 
\usepackage[T1]{fontenc} 
\usepackage[utf8]{inputenc} 
\usepackage{ISMIRlbd, cite, times, amsmath, url}
\usepackage{hyperref}
\usepackage{url}
\usepackage{float}
\usepackage{graphicx}
\usepackage{booktabs}
\usepackage{color}
\usepackage{circuitikz}
\usetikzlibrary{calc}

\title{Generative audio synthesis with a parametric model}

\threeauthors
 {Krishna Subramani} {IIT Bombay \\ {\tt \small krishna.subramani@iitb.ac.in}}
 {Alexandre D'Hooge} {ENS Paris-Saclay \\ {\tt \small dhooge@crans.org}}
 {Preeti Rao} {IIT Bombay \\ {\tt \small prao@ee.iitb.ac.in}}

 \def\authorname{Krishna Subramani, Alexandre D'Hooge, Preeti Rao}

\begin{document}

\maketitle

\begin{abstract}


Audio synthesis is of interest to people with different backgrounds such as music enthusiasts, performers, composers and researchers. With the advent of data-driven statistical modeling and abundant computing power, researchers are turning increasingly to deep learning for audio synthesis. Various approaches such as autoregressive modeling, GANs and VAEs have been proposed with varying degrees of success given the ultimate goal of modeling complex instrument sound sources, possibly while also achieving the flexible control of musically relevant attributes. 
Recently, Roche et al.\cite{roche_autoencoders_2018} studied different autoencoder architectures, like variational (VAE) and LSTM based autoencoders, to perform the frame-wise reconstruction of audio spectra. They also presented an analysis of the lower dimensional representation of the input spectrum, or ‘latent space’,  which a musician can explore to generate new sounds. Esling et al.\cite{esling_generative_2018} took this idea further and tried to incorporate structure into this latent space to match the perceptual timbre space of the corresponding instruments.  

In the interest of more flexible control over the generated sound however, it could be useful to somehow decompose the latent representation into relevant musical attributes such as pitch, dynamics and timbre. Such a possibility is more likely to be available with a parametric signal model rather than a general spectral representation such as the Fourier transform. Recognizing this in the context of speech synthesis, 
Blaauw et al.\cite{blaauw2016modeling} used a vocoder representation for speech, and then trained a VAE to model the frame-wise spectral envelope. Thus, by incorporating a powerful signal processing approach to separate the distinct perceptual attributes of the sound class, the network can be expected to learn a much more meaningful representation. 

\begin{figure}[ht]
\centering
    \begin{tikzpicture}[>=stealth, line width=.3mm, scale=0.95]
    \node[inner sep=0pt] (input) at (0,0) {\includegraphics[width=.2\textwidth]{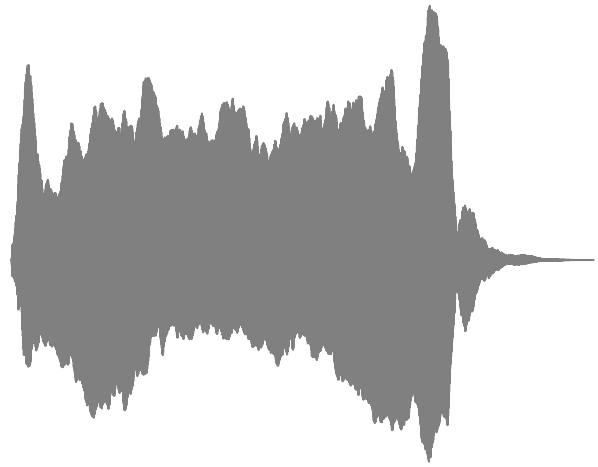}};
    \node[inner sep=0pt, draw] (fft) at ($ (input) + (4,0) $) {\includegraphics[width=.2\textwidth]{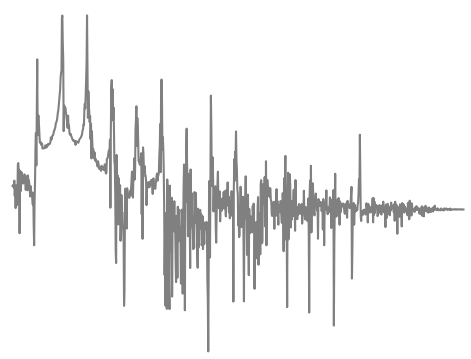}};
    \draw[] ($ (input) - (0,1) $) rectangle ($ (input) + (-.7,1) $)  node[above] {\hspace{2.5em} Sustain};
    \draw[->] ($ (input) + (-0,0) $) -- (fft) node [midway, above] {FFT};
    
    \node[draw, rotate=90] (enc1) at ($ (fft) + (3,0) $) {Encoder};
    \node[draw, rotate=90] (ls1) at ($ (enc1) + (1,0) $) {$Z$};
    \node[draw, rotate=90] (dec1) at ($ (ls1) + (1,0) $) {Decoder};
    \draw[dashed, rotate=90] ($ (enc1) + (-1, 0.5) $) rectangle ($ (dec1) + (1, -0.5) $) node[above] {\hspace{-9em}VAE};
    \draw[->] (enc1) -- (ls1);
    \draw[->] (ls1) -- (dec1);
    
    \draw[->] (fft) -- (enc1);
    \node[inner sep=0pt, draw] (out1) at ($ (dec1) + (2.5, 0) $) {\includegraphics[width=.2\textwidth]{./figs/fft.png}};
    \draw[->] (dec1) -- (out1);
    
    \node[inner sep=0pt, draw] (sf) at ($ (fft) + (-3, -3.7) $) {\includegraphics[width=.25\textwidth]{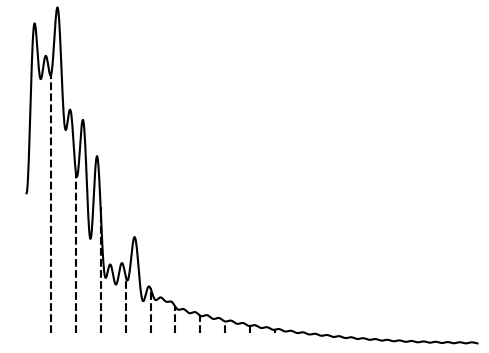}};
    
    \draw[->] (fft) -- (sf) node [below, xshift=1em, yshift=3.5em] {Source-Filter Model};
    
    \node[draw] (cc) at ($ (sf) + (3,0) $) {CCs};
    \draw[->] (sf) -- (cc);
    \node[draw] (f0) at ($ (sf) + (3,1.5) $) {$f_0$};
    \draw[->] (sf) -- (f0);
    \node[draw] (vel) at ($ (sf) + (3,-1.5) $) {Velocity};
    \draw[->] (sf) -- (vel);
    
    \node[draw, rotate=90] (enc2) at ($ (cc) + (2,0) $) {Encoder};
    \node[draw, rotate=90] (ls2) at ($ (enc2) + (1,0) $) {$Z$};
    \node[draw, rotate=90] (dec2) at ($ (ls2) + (1,0) $) {Decoder};
    \draw[dashed, rotate=90] ($ (enc2) + (-1, 0.5) $) rectangle ($ (dec2) + (1, -0.5) $) node[above] {\hspace{-9em}CVAE};
    \draw[->] (enc2) -- (ls2);
    \draw[->] (ls2) -- (dec2);
    
    \draw[->] (cc) -- (enc2);
    \draw[->] (f0) -- (enc2);
    \draw[->] (vel) -- (enc2);
    
    \draw[->] (f0) -| (dec2);
    \draw[->] (vel) -| (dec2);

    \node[inner sep=0pt, draw] (out2) at ($ (dec2) + (3, 0) $) {\includegraphics[width=.25\textwidth]{./figs/sf-model.png}};
    
    \draw[->] (dec2) -- (out2);
    
    \draw[dotted] ($ (fft) + (2, -1.5) $) rectangle ($ (out1) + (2, 1.5) $) node [above left] {\textit{Literature}};
    \draw[dotted] ($ (sf) + (-2.1, -2) $)  node [below right] {\textit{Proposed Work}} rectangle ($ (out2) + (2.1, 2) $);
    
    \end{tikzpicture}
    \caption{Flowchart of the state of the art audio synthesis pipeline (upper branch) and our proposed work (lower branch). $Z$ represents the latent space learned by the (C)VAE.}
    \label{model}
\end{figure}

The parametric representation we adopt is inspired from the analysis pipeline of \cite{caetano_musical_2013}, where ``perceptually motivated'' representations of audio are proposed for sound morphing. A Source Filter decomposition is applied to the harmonic component of the spectrum extracted by the Harmonic model \cite{serra1997musical}. The filter is estimated as the envelope of the harmonic spectrum and represented via low-dimensional cepstral coefficients. 
Therefore, as opposed to simply training a VAE on the full magnitude spectrum (upper branch in \autoref{model}), we train a conditional VAE (CVAE) on the real cepstral coefficients (CCs) conditioned on the pitch and velocity (lower branch in \autoref{model}) assuming that a suitably labeled dataset is available for the supervised training of the network. Our motivation to use a CVAE comes from the fact that the Source and Filter are not entirely decoupled for musical instruments. 
By conditioning on the pitch and velocity, our model is expected to generate the spectral envelope more accurately. 


We use a subset of the NSynth \cite{engel_neural_2017} dataset in our work. We have implemented the parametric representation and used it to successfully train a CVAE network. 
Trained on two different instruments, the network is capable of generating new sounds with a hybrid timbre. We have been able to test the validity of our model by training on sounds restricted to the odd MIDI pitches and examining the generation of the sounds with the missing pitches. The network indeed learned how to fill in the missing pitches, returning sounds with the expected timbre. Furthermore, the network could even produce perceptually acceptable continuous sounds when conditioned with a semi-continuous pitch frequency sweep (one octave divided into small steps). Finally, we observed that using a parametric representation allows us to use a very small neural network that can be trained in under an hour (on a mobile Nvidia GeForce GTX 1050Ti). 


Our network is currently trained only on the sustained segments of selected instrument sounds. By applying instrument-specific temporal segmentation, we plan to augment the network toward generating realistic instrument sounds with the components of attack, sustain and decay. Moreover, we are only training on the harmonic part of sounds while the residual also contains important information identifying the instrument (consider the bowing sound of a violin). Future work shall address this issue either by modifying the current CVAE or by using two networks that would exchange information during synthesis to produce perceptually better sounds.

\end{abstract}


\begin{acknowledgments}
The authors thank Prof. Xavier Serra for insightful discussions on the problem. 
\end{acknowledgments}

\bibliography{ISMIRtemplate}

\begin{thebibliography}{1}

\bibitem{blaauw2016modeling}
Merlijn Blaauw and Jordi Bonada.
\newblock Modeling and transforming speech using variational autoencoders.
\newblock In {\em Interspeech}, pages 1770--1774, 2016.

\bibitem{caetano_musical_2013}
Marcelo Caetano and Xavier Rodet.
\newblock Musical {Instrument} {Sound} {Morphing} {Guided} by {Perceptually}
  {Motivated} {Features}.
\newblock {\em IEEE Transactions on Audio, Speech and Language Processing},
  21(8):1666--1675, 2013.

\bibitem{engel_neural_2017}
Jesse Engel, Cinjon Resnick, Adam Roberts, Sander Dieleman, Douglas Eck, Karen
  Simonyan, and Mohammad Norouzi.
\newblock Neural {Audio} {Synthesis} of {Musical} {Notes} with {WaveNet}
  {Autoencoders}.
\newblock {\em arXiv:1704.01279 [cs]}, April 2017.
\newblock arXiv: 1704.01279.

\bibitem{esling_generative_2018}
Philippe Esling, Axel Chemla-Romeu-Santos, and Adrien Bitton.
\newblock Generative timbre spaces: regularizing variational auto-encoders with
  perceptual metrics.
\newblock {\em arXiv:1805.08501 [cs, eess]}, May 2018.
\newblock arXiv: 1805.08501.

\bibitem{roche_autoencoders_2018}
Fanny Roche, Thomas Hueber, Samuel Limier, and Laurent Girin.
\newblock Autoencoders for music sound modeling: a comparison of linear,
  shallow, deep, recurrent and variational models.
\newblock {\em arXiv:1806.04096 [cs, eess]}, June 2018.
\newblock arXiv: 1806.04096.

\bibitem{serra1997musical}
Xavier Serra et~al.
\newblock Musical sound modeling with sinusoids plus noise.
\newblock {\em Musical signal processing}, pages 91--122, 1997.

\end{thebibliography}

\end{document}